# Probing spin dynamics and quantum relaxation in LiY$_{0.998}$Ho$_{0.002}$F$_4$ via $^{19}$F NMR

M. J. Graf [1,2], A. Lascialfari[2], F. Borsa[2], A. M. Tkachuk[3], and B. Barbara[4]

[1] Department of Physics, Boston College, Chestnut Hill, MA 02467 USA
[2] Dipartimento di Fisica "A. Volta" e Unità INFM di Pavia, Via Bassi 6, I27100 Pavia, Italy
[3] All-Russia Scientific Center "S. I. Vavilov State Optical Institute," 199034 St. Petersburg, Russia
[4] Laboratoire de Magnétisme Louis Néel, CNRS, BP166, 38042 Grenoble Cedex-09, France

## Abstract

We report measurements of $^{19}$F nuclear spin-lattice relaxation $1/T_1$ as a function of temperature and external magnetic field in LiY$_{0.998}$Ho$_{0.002}$F$_4$ single crystal, a single-ion magnet exhibiting interesting quantum effects. The $^{19}$F $1/T_1$ is found to depend on the coupling with the diluted rare-earth (RE) moments. Depending on the temperature range, a fast spin diffusion regime or a diffusion limited regime is encountered. In both cases we find it possible to use the $^{19}$F nucleus as a probe of the rare-earth spin dynamics. The results for $1/T_1$ show a behavior similar to that observed in molecular nanomagnets, a result which we attribute to the discreteness of the energy levels in both cases. At intermediate temperatures the lifetime broadening of the crystal field split RE magnetic levels follows a $T^3$ power law. At low temperature the field dependence of $1/T_1$ shows peaks in correspondence to the critical magnetic fields for energy level crossings (LC). The results can be explained by inelastic scattering between the fluorine nuclear spins and the RE magnetic levels. A key result of this study is that the broadening of the levels at LC is found to be become extremely small at low temperatures, about 1.7 mT, a value which is comparable to the weak dipolar fields at the RE lattice positions. Thus, unlike the molecular magnets, decoherence effects are strongly suppressed, and it may be possible to measure directly the level repulsions at avoided level crossings.

## Introduction

The material $LiY_{1-x}Ho_xF_4$ has proven to be very important in the studies of magnetism. Pure $LiHoF_4$ (x = 1) exhibits dipolar coupling between the magnetic $Ho^{3+}$ ions, and becomes a ferromagnet below a Curie temperature of $T_C$ = 1.5 K.[1] Dilution of Ho with non-magnetic Y lowers $T_C$, and for 0.1 < x < 0.5 the low temperature ground state is a spin-glass.[2] For somewhat lower concentrations (x ≈ 0.05), the ground state appears to be an unusual quantum "anti glass".[3,4] Two important features have made this system particularly useful for studying the evolution from correlated ferromagnet towards single-ion behavior: (1) The material is essentially isostructural over the entire range of x, allowing for a systematic study of dilution of the magnetic moments; and (2) Crystal field effects split the J = 8 ground state manifold ($g_J$ = 5/4), resulting in a new ground state that is an Ising-like doublet and is separated from the singlet first excited state by roughly 10 K[5] (see Fig. 1a). From the exact diagonalization of the Hamiltonian it is found that the ground state doublet corresponds to an average $\langle J_z \rangle \cong 5.3$, or an effective g factor $g_{eff.}$ = 13.3, as determined from EPR spectroscopy.[6] Thus at low temperatures a single $Ho^{3+}$ ion behaves as an atomic magnet with a frozen magnetic moment due to the large axial crystal field anisotropy along the tetragonal c-axis, which is the quantization z-axis. The energy barrier for reorientation, $E_b \approx$ 10 K, is determined by the separation to the first excited electronic singlet (see Fig.1a). This situation allows one to extend the study of quantum relaxation and tunneling from large spin molecules such as Mn12 [7] and Fe8 [8] to single rare-earth moments, as recently shown in a crystal of $LiY_{1-x}Ho_xF_4$ with very dilute Ho concentrations (x ≈ 0.002).[9] Indeed, as shown in Ref. 9, the quantum dynamics of single ion rare-earth magnets is made particularly interesting by the coupling of the electronic moment with the nuclear moment. The hyperfine coupling to the $^{165}$Ho nuclear spin (I = 7/2), i.e. $A_J$ **J·I,** where $A_J$ is the hyperfine constant, further splits the ground state Ising doublet into eight two-fold degenerate levels equally spaced by $\Delta E = \langle J_z \rangle A_J \approx$ 205 mK. Application of a small magnetic field along the crystalline c-axis introduces Zeeman splittings of these degenerate levels. Calculations predict that level crossings (and some avoided level crossings) occur at longitudinal fields given by $H_n = nA_J / 2g_J \mu_B \approx n$ 23 mT, where n is an integer with $-7 \leq n \leq 7$, as shown in Fig. 1b. This has been confirmed by the observation of quantum tunneling of the magnetization (QTM), as evidenced by steps in the magnetization at the resonant field values,[9] in analogy with the phenomenon observed in high-spin molecular magnets Mn12 and Fe8. Much smaller features are observed at half-integral values[10,11] and correspond to two-ion cooperative tunneling processes.

Nuclear magnetic resonance (NMR) has proven to be a useful tool in probing the spin dynamics associated with level crossings and QTM.[12-15] Performing NMR measurements on the $^{165}$Ho nucleus would be extremely difficult due to the very large nuclear quadrupolar moment (Q ≈ 2 barn) which is likely to complicate the nuclear energy level scheme in a non cubic crystal environment. Thus we chose to use the $^{19}$F nucleus (I = 1/2) as a probe. Although the many fluorine nuclei are only weakly coupled by dipolar hyperfine interactions with the diluted $Ho^{3+}$ moments, we found that the coupling is sufficiently strong to dominate the $^{19}$F spin-lattice relaxation, thus allowing us to obtain information about the $Ho^{3+}$ spin dynamics.

**Experimental**

We have made measurements on a single crystal sample of $LiY_{0.998}Ho_{0.002}F_4$, as well as pure $LiYF_4$ for reference. NMR data were taken in a gas flow cryostat and pumped liquid helium bath for data above and below 4 K, respectively. NMR spectra were obtained using a TecMag Apollo spectrometer, with static magnetic fields provided by an electromagnet. The samples were oriented such that the c-axis formed an angle of approximately 30 degrees with the static field. The component of the field parallel to the sample c-axis determines the resonant tunneling values for QTM, while the perpendicular component breaks the uniaxial symmetry and so in principle increases the rate of QTM,[9] albeit by an unknown amount. This tilted configuration allowed us to work at manageable Larmor frequencies (ν > 4 MHz) as determined by the total magnetic field strength while probing the low fields where QTM is known to occur ($H_z$ ≤ 161 mT). A π/2-π/2 sequence was found to produce a maximum spin echo. Fourier transforms of the spectra yielded frequencies consistent with the $^{19}$F gyromagnetic ratio of 40.05 MHz/T; no attempts were made to precisely measure the chemical shift. Typical driving pulses were 1-3 μs in duration, while the $^{19}$F linewidth was 40 kHz FWHM. The spin-spin relaxation time $T_2$ was determined to be approximately 25 μs and temperature independent from the decay of the spin-echo signal. A comparable value for $T_2^*$ (≈ 15 μs) was extracted from the free-induction decay curves, indicating that the inhomogeneous broadening due to the coupling with the $Ho^{3+}$ is less than the homogeneous nuclear dipolar broadening i.e. $T_2^* \leq T_2$. The source of homogeneous broadening is due to the $^{19}$F-$^{19}$F homonuclear dipolar interaction which is estimated to be $(\Delta\omega^2)^{1/2} \approx 90$ KHz-rad from the Van Vleck second moment formula.[16] The inhomogeneous broadening originates from the smaller heteronuclear $^{19}$F-$^{7}$Li and $^{19}$F-$^{89}$Y dipolar interaction. The nuclear-nuclear and nuclear-electron dipolar interaction of the $^{19}$F nucleus with the $^{165}$Ho nucleus and the Ho$^{3+}$ electron moment respectively are negligible due to the small concentration of rare-earth magnetic ions.

The short spin-lattice relaxation time $T_1$ (milliseconds) in LiY$_{0.998}$Ho$_{0.002}$F$_4$ was determined by monitoring the nuclear magnetization following a saturating sequence of $\pi/2$ pulses. We have also made limited reference measurements in undoped LiYF$_4$. Due to the very long spin-lattice relaxation time (tens of seconds) in this case the exponential variation of the spin echo signal with repetition time was used to determine $T_1$.

**Results**

At high temperatures (T ≥ 20 K), when the $T_1$ is longer, the recovery of the nuclear magnetization for LiY$_{0.998}$Ho$_{0.002}$F$_4$ was found to be almost exponential, indicating that a common $^{19}$F spin temperature could be achieved via nuclear spin diffusion over times much shorter than the typical spin-lattice relaxation time. In the vicinity of 8 – 20 K, the relaxation rate increases significantly, and the recovery of the nuclear magnetization shows an initial part at short times which varies as the square root of time, followed by an exponential tail at long times. This behavior is associated with a spin-diffusion-limited relaxation regime.[17] This behavior is shown in Fig. 2. The curves for 12 K, where $T_1$ has become very short, exhibit behavior as described above, and is to be contrasted to the nearly pure exponential decay observed at 40 K, where $T_1$ is longer. In the diffusion-limited regime we extract the relaxation rate from the long time behavior. Henceforth we will present the data in terms of a single parameter, $1/T_1$, although one should be aware that the precise relation between the relaxation rate and the spin dynamics may be different in the different regimes, as discussed later on.

In Figure 3 we show $1/T_1$ for LiY$_{0.998}$Ho$_{0.002}$F$_4$ versus temperature for Larmor frequencies of 5.35 MHz, 10.30 MHz, and 30.10 MHz, corresponding to magnetic field values of 134 mT, 331 mT, and 751 mT, respectively. Two measurements were performed on pure LiYF$_4$ at a frequency of 10.30 MHz and yielded $1/T_1 = 5.9 \times 10^{-5}$ (msec$^{-1}$) at 10 K and $3.1 \times 10^{-4}$ (msec$^{-1}$) at 20 K. As can be seen, the addition of a very small amount of Ho has increased the $^{19}$F relaxation rate by three orders of magnitude. This unambiguously shows that the $^{19}$F nuclear relaxation is driven by the fluctuating hyperfine field of the $Ho^{3+}$ moments. Since the $Ho^{3+}$ moments are highly diluted only the $^{19}F$ nuclei nearby are relaxed directly by the electronic moments while the remaining fluorine nuclei are brought to the same spin temperature by a $T_2$ process. In any case the effective measured relaxation rate is directly related to the $Ho^{3+}$ spin dynamics. The focus of this study is the temperature and magnetic field dependence of the $Ho^{3+}$ spin dynamics as probed by the effective $^{19}F$ relaxation rate.

In the range 10 - 15 K $1/T_1$ exhibits a maximum that increases with size and shifts to decreasing temperatures with decreasing frequency. This behavior is very similar to the one observed in a variety of antiferromagnetic molecular rings[12,18] and in single molecule magnets,[19] and will be discussed in the next section.

To look for phenomena associated with the energy level crossings (Fig. 1b), as reported in Refs. 7-11, we measured the magnetic field dependence of $1/T_1$ at a temperature of 1.7 K, a temperature where thermally activated QTM was observed via bulk magnetization measurements of $LiY_{0.998}Ho_{0.002}F_4$. The sample magnetization is hysteretic, so to minimize the sample's history dependence we took measurements by ramping the magnetic field to a value above 200 mT, then measuring $1/T_1$ with decreasing magnetic field. As shown in Fig. 4a, we observe a series of peaks in $1/T_1$ which we associate with the level crossings shown in Fig. 1b. The exact sample orientation is not known, so we assume that the highest field maximum in $1/T_1$ occurs at a field value of 161 mT, corresponding to the n = 7 resonant tunneling value shown in Fig. 1b. This is observed to occur when our total magnetic field strength is 187 mT. Thus we calculate the orientation of the sample's c-axis with respect to the magnetic field to be 31 degrees, very close to our visual estimate of the orientation. The energy level diagram shown in Fig. 1b is dependent on the component of the field along the z-quantization axis (sample c-axis), so in Fig. 4b we plot our measured $1/T_1$, minus a background contribution discussed in the next section, versus $H_z = H \cos 31^o$. The locations of the observed peaks are in very good agreement with the expected locations for n = 4, 5, 6, and 7 as determined from magnetization measurements, with an average spacing of 23 mT. A scan of $1/T_1$ versus H at 4.2 K in the vicinity of n = 5 shows a very broad peak which is difficult to extract from the increased background value of $1/T_1$. In Fig 5 we show the temperature variation of $1/T_1$ both at a field value corresponding to the n = 5 peak and at a field value off the peak and intermediate between n = 5 and n = 6.

## Discussion

There are two important aspects of the spin dynamics and quantum relaxation of the impurity rare earth ion which are probed by the $^{19}$F nuclear spin lattice relaxation rate. The first one is the spin dynamics at intermediate temperatures, namely temperatures of the order of the gap between the ground state Ising doublet and the first singlet excited state (about 10 K in Fig.1a). The second one is the quantum tunneling regime which occurs at low temperature in the vicinity of the critical field corresponding to the level crossings within the Ising doublet split by the hyperfine interaction with the $^{165}$Ho nucleus, as shown in Fig.1b. We will discuss the two temperature regions separately in the following subsections.

### a) Spin dynamics at intermediate temperatures

The nuclear spin-lattice relaxation of $^{19}$F is due to the dipolar coupling of the nucleus with the magnetic moment of the rare-earth paramagnetic moment as proved by the fact that in pure LiYF$_4$ the $^{19}$F $1/T_1$ is three orders of magnitude longer than in the Ho doped sample. Thus, in the weak collision perturbative approach the transition probability of a nucleus at distance r from the impurity is given by[16]

$$P = A\, g(\theta)/r^6\ f(\omega_N)\ , \qquad (1)$$

where the first three factors represent the dipolar coupling with θ defined as the angle the line joining the nucleus and the rare-earth ion forms with the magnetic field, and f(ω$_N$) is the electronic spin spectral function evaluated at the nuclear Larmor frequency. In order to give an explicit form to f(ω) one can refer to one of two limiting cases: (i) one can assume a semiclassical model in which the S$_z$ component of the electron spin is treated as a classical quantity and the spectral function f (ω) is evaluated as the Fourier transform of the correlation function of S$_z$(t), or (ii) one can use a quantum mechanical approach whereby the transition probability is calculated in terms of the matrix elements of the S$_z$, S$_\pm$ operators between the total magnetic quantum states of the rare-earth ion. The first case is utilized whenever the spectrum of the magnetic states is a quasi-continuum and the temperature is sufficiently high to populate a large number of states. The second approach should be applied when the magnetic levels are split by crystal field effects and are well separated in energy so that at the working temperature only a limited number of them are occupied. This is the case of the rare-earth $Ho^{3+}$ ion in $LiYF_4$ below about 50 K. It should be mentioned that another system where only discrete and well separated magnetic energy levels are present is the finite size molecular nanomagnets.[12] Thus one expects many similarities in the NMR response of these two systems.

Going back to the explicit form of $f(\omega)$ in Eq. 1 one can refer to the discussion for the case of molecular nanomagnets. It has been shown[20,21] that since the energy separation between magnetic states is much larger than the nuclear Zeeman splitting, the only energy conserving matrix elements involve the $S_z$ component of the electron spin, and $f(\omega)$ is ideally a delta function centered at zero frequency. In practice, the delta function will be broadened by interactions with the environment so that $f(\omega)$ is non-zero at $\omega = \omega_N$, allowing relaxation to occur. One then obtains quasi-elastic terms in the nuclear relaxation described by Eq. 1 with $f(\omega_N)$ represented phenomenologically by a Lorenztian (or Gaussian) function of width $\Gamma$ :

$$P = A\, g(\theta) / r^6 \ \Gamma / (\Gamma^2 + \omega_N^2) \ . \qquad (2)$$

It should be noted that Eq. 2 is formally identical to the expression obtained in a semi-classical approach under the assumption of an exponential correlation function for the random variable $S_z(t)$ described by a correlation time. In our case, however, the parameter $\Gamma$ represents the lifetime broadening of the rare-earth magnetic states due to the spin-phonon relaxation transition probability.

Once the expression for the transition probability for the nucleus has been established (Eq. 2) one can then address the problem of the total nuclear magnetization relaxing towards equilibrium. Since the relaxation centers are very diluted one has to consider the combined effect of relaxation of nuclei near the rare-earth ions and the nuclear spin diffusion.[17] If the nuclear Zeeman energy can diffuse to the paramagnetic ion via a $T_2$ process faster than the paramagnetic ion can transmit it to the lattice by a $T_1$ process (rapid diffusion) one observes an exponential decay of the nuclear magnetization. If, on the other hand, the nuclear spin diffusion is not fast enough (diffusion limited case) one should observe an initial transient recovery given by a $t^{1/2}$ law followed at long time by an exponential recovery. The different regimes can be identified according to Blumberg[17] on the basis of the values of parameters based on : (i) the nuclear $T_2$ which is here about 25 μsec, (ii) the average distance between $Ho^{3+}$ ions i.e., $R_v = 3.3$ nm (determined by the Ho concentration, and the tetragonal unit cell with dimensions c = 1.074 nm and c = 0.5175 nm,[22] which contains four possible Ho sites), and (iii) the spin-lattice relaxation time, which varies from 2 to 50 msec in this study. At temperatures well above the peak in Fig. 3a the conditions for rapid diffusion are met and one observes an exponential decay of the nuclear magnetization. At temperatures corresponding to the region of the peak in $1/T_1$ an initial fast decay is observed indicating the presence of a diffusion limited relaxation. However, by excluding the decay at short time of the magnetization one can extract the long-time relaxation rate which is proportional to the spectral function $f(\omega_N)$ and which defines the spin dynamics. Thus, we assume that the data in Fig.3a are described by the simple expression

$$1/T_1 = C \, \Gamma / (\Gamma^2 + \omega_N^2) \quad , \qquad (3)$$

where the constant C is obtained from integration of Eq. 2 over all nuclei and so contains both the average strength of the nuclear electron dipolar interaction and the nuclear spin diffusion constant.[16] If one assumes that the lifetime broadening of the magnetic states is described by a power law $\Gamma \propto T^\alpha$ it follows from Eq. 3 that the relaxation rates rescaled to the peak value should follow the simple expression[18]

$$[1/T_1(H,T)]/[1/T_1(H,T_0)] = 2 \, t^\alpha / (1+t^{2\alpha}) \, , \qquad (4)$$

where $t = T/T_0(H)$ and $T_0(H)$ is the temperature at which the relaxation peak occurs at a given field magnetic field H (see Fig.3a). In the work of Reference 18, the data were renormalized by a factor of $\chi(T)T$ in order to account for the temperature dependent magnetic fluctuations. This is not required here, as we have measured the magnetic susceptibility (H along c-axis) and find that it obeys Curie's Law in the temperature range of interest, thus yielding a constant value for $\chi(T)T$ which is incorporated in the fitting constant C in Eq. 3. The data, rescaled according to Eq. 4, are shown in Fig. 3b. As can be seen the fit in the region of the peak is very good with the choice of $\alpha = 3.0 \pm 0.3$ as the only fitting parameter in the formula. This result is very similar to the one found in antiferromagnetic molecular rings.[18] This is not too surprising since in both cases the nuclear relaxation proceeds via a direct quasi-elastic process within a given magnetic state broadened by spin-phonon interactions. A key feature of Eq. 3 is that $1/T_1(H,T)$ has a local maximum for fixed H and variable T, occurring at a temperature $T_0(H)$, at which $\Gamma(T)$ equals $\omega_N$. This implies that the lifetime broadening becomes very small as the temperature is lowered. By using the data in Fig.3a and setting $\Gamma(T_0) = \omega_N$ one can estimate the absolute value for the lifetime broadening i.e., $\Gamma = 5.0 (\pm 2.0) 10^4 T^3$ (Hz rad). This value is close to the one found in the antiferromagnetic ring Cr8.[18] This remarkable finding can be partially explained by the fact that the two systems have comparable values for the energy gap between the ground state and first excited states. For Cr8 the gap is about 8.6 K from the singlet ground state of total S=0 to the first triplet excited state S=1,[13] while for $Ho^{3+}$ in $LiYF_4$ the gap between the Ising doublet and the first excited singlet state is about 10 K, as shown in Fig.1. This observation may imply that at low temperatures the value of the spin-phonon lifetime broadening is mainly determined by density of acoustic phonon states at the energy corresponding to the gap from the ground state to first excited state. The simple model resulting in Eq. 4 subsumes all the system details into a single, field-independent parameter, $\Gamma$. This is sufficient to understand the underlying physics of the lifetime broadening. However, the model does not include the magnetic field dependence of the energy levels (see Fig. 1), and so is not adequate for detailed fits of the individual curves. $\Gamma$ will in fact be field-dependent.

An important consequence of the present finding is that the lifetime broadening of the $Ho^{3+}$ magnetic states becomes vanishingly small at low temperatures. Since the other source of broadening is the dipolar interaction, which is very small because of the low concentration of paramagnetic moments, one expects strongly reduced decoherence effects in the quantum dynamics at the level crossings. This opens up the possibility of directly measuring the magnitude of the level repulsion, which can be interpreted as a tunneling splitting.

### b) Spin dynamics at level crossings

At low temperature (T = 1.7 K) the $^{19}F$ relaxation rate as a function of the external magnetic field shows four peaks in correspondence to the level crossings n = 4, 5, 6 and 7 (see Fig.4a). As shown in the inset of Fig.4a, a linear fit of the peak position vs. n yields $H_n = n$ 27 mT. By considering that our sample is aligned so as to form an angle of about $31^0$ with the anisotropy z axis, the longitudinal component of the field is $H_z = H \cos 31^o$. Then one has $H_{zn} = n$ 23mT, in good agreement with magnetic susceptibility measurements.[9]

The peaks arise on the top of a background relaxation which decreases with increasing field H. The background relaxation can be fitted with $1/T_1 = 0.96 / H^{0.8}$ msec$^{-1}$, with H in mT (see Fig.4a). The background relaxation should originate from the same quasi-elastic contribution described by Eq. 3, which appear to fit the higher temperature data. This is to be contrasted with the inelastic contribution which dominates near the critical field for level crossing as discussed below. The interpretation of the field dependence of the background relaxation is not clear at present. From the Eq. 3 one would expect that at low temperature where $\Gamma << \omega_N$ the nuclear relaxation rate should go as $1/T_1 \propto 1/\omega_N^2 \propto 1/H^2$. On the other hand since at low temperature the time-recovery of the nuclear magnetization may be governed by diffusion limited processes, one may have a situation where the long time behavior of the recovery curve is no longer given by $1/T_1 \propto f(\omega_N)$. For example, we note that the observed $1/T_1 \propto H^{-0.8}$ behavior is close to that expected in the diffusion limited regime under certain limiting conditions, for which the exact solution of the diffusion equation yields for the long time recovery of the nuclear magnetization[17] $1/T_1 \propto f(\omega_N)^{1/4}$ and thus $1/T_1 \propto \omega_N^{-0.5} \propto H^{-0.5}$. In any case, the correct interpretation of the field dependence of the background relaxation is not relevant here since we use the fitting curve for the sole purpose of subtracting the background relaxation from the total relaxation in order to isolate the inelastic contribution dominating near the energy level crossings.

In order to analyze the peaks in the relaxation rate at the critical fields for level crossing we have replotted the data in Fig.4b. The background relaxation has been subtracted and the data are plotted as a function of the z-axis component of the magnetic field. In the proximity of the critical field for level crossing the transverse hyperfine contribution and the transverse component of the applied field, given for our sample orientation by $H_T = H \sin 31^0$, can induce transitions between the two eigenstates $|\psi_1^- I_{z1}\rangle$ and $|\psi_1^+ I_{z2}\rangle$, where $\psi_1^\pm$ refer to the ground state Ising doublet and $I_z$ refers to the $^{165}$Ho nuclear spin. Since the energy states near the level crossings are very close in energy it is possible to have inelastic scattering of the $^{19}$F nuclear spin by the magnetic states at the Larmor frequency $\omega_N$. This inelastic contribution to relaxation has been discussed in the framework of Moriya theory[23] for magnetic relaxation and shown to be describable by an expression of the form [21]

$$1/T_1 = B^2 \ W / [W^2 + (\omega_N - \Delta)^2] , \qquad (5)$$

where W is the width of the magnetic states and $\Delta$ is the energy separation in frequency units between the two magnetic states at level crossing. In the present case the energy separation can be expressed by $\Delta = [1.3 \times 10^{18} (H - n\, 23)^2 + \Delta_n^2]^{1/2}$ (Hz rad /mT) for the $n^{th}$ LC. The parameter $\Delta_n$ represents the hyperfine induced tunnel splitting at the $n^{th}$ (avoided) level crossing. The data in Fig.4b have been fitted with Eq. 5. In the fitting procedure we have treated B as a T- and H-independent fitting constant since it is related to the hyperfine dipolar coupling of the $^{19}$F nucleus to the $Ho^{3+}$ ion. Furthermore, for each n one should consider as many terms of the form of Eq. 5 as there are level crossings (or avoided level crossings), e.g., four for n = 4, three for n = 5, two for n = 6 and one for n = 7 (refer to Fig.1). For simplicity we assume each level crossing has the same value of $\Delta_n$. Each term should be weighted by a Boltzmann factor $\exp(-E_n/k_BT)/Z$ where Z is the partition function and $E_n/k_B = 0.205\, n$ (Kelvin) for the lowest energy LC at n. When more than one LC is present at the same field H, each higher energy LC is separated by 205 mK from the lower one (see Fig.1). The effect of the different thermal populations of the levels at each LC explains qualitatively the decrease of intensity of the peaks for increasing critical field.

The central result here is the fact that the peaks in $1/T_1$ are very narrow, namely only a few mT wide. The width of the peaks can be either due to the intrinsic broadening W of the magnetic levels of the $Ho^{3+}$ ions, or to the tunnel splitting $\Delta_n$, or both. In fact the data can be fitted well either by assuming $W = \Delta_n = 2 \times 10^9$ (Hz rad) (equivalent to 1.72 mT and 15.2 mK, in magnetic and thermal units, respectively, using $g_{eff} = 13.3$ for the electronic levels in the conversion of units), or by assuming one of the two parameters is given by the above value and the other one is much smaller. In order to eliminate the ambiguity one can resort to comparison with other experimental results and/or theoretical estimates. The source for the broadening W at such low temperatures should originate entirely from the dipolar interaction among the $Ho^{3+}$ magnetic moments. An estimate of the dipolar field at the average distance between two rare-earth moments yields a value of 0.6 mT, or 5.3 mK. On the other hand, an estimate of the tunnel splitting at some of the avoided level crossings from magnetization measurements[9] yields a value of 25 mK. Both these estimates are comparable to the observed broadening. We note that our data show that the peak broadening is approximately independent of magnetic field. The two parameters W and $\Delta_n$ are expected to have a different field dependence. In fact, W should be weakly field dependent since at low temperature the $Ho^{3+}$ magnetic moment is frozen, while the tunneling splitting $\Delta_n$ typically increases strongly with the applied transverse field, although no exact calculations of the effect are currently available.. Thus our results seem to suggest that the level broadening is primarily determined by W. However, this conclusion is only tentative, and more detailed measurements of the evolution of the width of the nuclear relaxation peaks with transverse magnetic field component and with the Ho$^{3+}$ concentration are required in order to independently determine W and $\Delta_n$. It is noted that in the presence of a pure level crossing, i.e. $\Delta_n$ = 0, Eq. 5 would predict the observation of two peaks, one when the gap $\Delta$ becomes equal to $\omega_N$ from above and one from below the LC condition. However, the two peaks are separated by 0.1 mT and would thus be unobservable even with a width W of only 1 mT.

Finally we consider the temperature dependence of the relaxation rate at constant field, which is shown in Fig. 5. One set of data was taken at the critical field associated with the n = 5 level crossing (i.e., at a peak) while the second set of data was taken at a field intermediate (off the peak) between the n = 5 and n = 6 LC's. The observed temperature dependence has been fitted by the sum of two contributions. The first one is the quasi-elastic contribution which should behave as T$^3$ while the second contribution is the inelastic contribution present near the LC conditions. At the level crossing the inelastic contribution should be proportional to the population of the magnetic levels at the LC and its temperature dependence should be simply described by a sum of Boltzmann factors. Thus the data at the LC were fitted with the expression

$$1/T_1 = E\, T^3 + G\, (\Sigma_i\, G_i\, \exp(-E_i/k_B T)\,/Z)\,, \qquad (6)$$

where the sum is over all the i levels which have level crossings at the magnetic field associated with n = 5 (115 mT), and $G_i$ is the degeneracy of the i$^{th}$ energy level (that is, 2). Z is the partition function evaluated at that field value. As shown in Fig.5 a good fit for the data at the n = 5 LC is obtained with E = 0.95 (Hz rad/K$^3$) and G = 92 (Hz rad), values that are compatible with the separate fits of the quasi-elastic contribution in Fig.3 and of the inelastic contribution in Fig.4. The data at a magnetic field in between the two LC's n = 5 and n = 6 can also be fitted with the same elastic contribution ($\propto$ T$^3$), and an inelastic contribution which has been reduced by a factor of two. Both fits are shown in the figure as solid lines. However, the quality of the fit is rather insensitive to the values of the fitting parameters over this limited temperature range. We conclude that the observed temperature dependence is consistent with the analysis shown in Figs. 3 and 4. More data, taken to lower temperatures (T < 1 K) where 1/T$_1$ is expected to decrease rapidly towards zero, are required to confirm the efficacy of Eq. 6 and accurately determine the values of the fitting parameters.

## Conclusions

To summarize, we present data for the $^{19}$F spin-lattice relaxation rate $1/T_1$ in LiY$_{0.998}$Ho$_{0.002}$F$_4$, and find that it is an excellent probe of the spin dynamics of the Ho$^{3+}$ ion. This behavior is well described by a model which has the $^{19}$F nuclei approaching equilibrium via spin diffusion, and those near the Ho centers are then relaxed to the lattice. At intermediate temperatures (4 K < T < 50 K) $1/T_1$ is determined by the discrete energy spectrum of the Ho ion, exhibiting behavior which is comparable to molecular magnets with similar energy level structures (e.g., Cr8). The quasi-elastic lifetime broadening is found to vary like $T^3$, and is believed to result from spin-phonon interactions. An important consequence of this behavior is that the low-temperature broadening of the magnetic energy levels will be very small, due primarily to the weak dipolar fields. Decoherence effects are suppressed, and the resulting long lifetimes are experimentally confirmed by our observation of very sharp peaks at magnetic field induced Ho$^{3+}$ level crossings. Thus, this system is an excellent candidate to allow for a direct measurement of the level repulsion / tunnel splitting at level crossings through detailed studies of the variation of the peak broadening with transverse magnetic field and Ho concentration. Such a study is now underway.

## Acknowledgements

MJG would like to thank his colleagues at the University of Pavia for their hospitality during his sabbatical visit. This work was supported by INTAS grant 03-51-4953, and the QUEMOLNA and MAGMANET programs.

[21]   A. Cornia, A. Fort, M.G. Pini and A. Rettori, Europhysics Lett. **50**, 88 (2000).

[22]   E. R. Thoma et al., J. Phys. Chem. **65**, 1906 (1961)

[23]   T. Moriya, Prog. Theor. Phys. **16**, 23 (1956); see also ibid. **28**, 371, (1962).15

**Figure Captions**

Figure 1. Variation of the low-lying energy states with magnetic field. (a) Levels calculated for $Ho^{3+}$ ion (J = 8) with crystal field splitting. (b) Close-up view of the ground state doublet, which strongly couples to the I = 7/2 Ho nucleus, resulting in 16 electro-nuclear states (from Ref. 9).

Figure 2. (a) Time decay of the sample magnetization (normalized to 1 at t = 0) at 12 K and 40 K for $LiY_{0.998}Ho_{0.002}F_4$ taken at a frequency of 13.30 MHz. Solid lines are the long-time exponential fits to the data, from which we extract $1/T_1$. (b) Plot of the recovery of the longitudinal magnetization versus the square root of time at the same two temperatures. Solid lines are guides to the eye.

Figure 3. (a) Temperature variation of the $^{19}F$ spin-lattice relaxation rate at three frequencies. (b) Data in Fig. 3a, normalized to a maximum value of 1, as a function of $t = T/T_0$, where $T_0$ is the temperature at which the peak occurs in part (a). The solid line is a fit to Eq. 3 with $\alpha = 3.0$.

Figure 4. (a) Variation of the $1/T_1$ of $^{19}F$ in $LiY_{0.998}Ho_{0.002}F_4$ with the applied magnetic field at low field values. The solid curve is a fit of the background relaxation according to $0.96 / H^{0.8}$ (msec$^{-1}$). The inset shows the peak magnetic field value plotted vs. the index n of the level crossing. (b) Plot of the relaxation rate with the background term subtracted vs. the component of the applied field in the direction of the anisotropy axis z. The full curves are fit of the data according to the inelastic contribution as discussed in the text.

Figure 5. Temperature dependence of the $^{19}F$ relaxation rate at a field value corresponding to the energy level crossing at n = 5, and to a value away from the peak between n = 5 and n = 6. The solid curves correspond to a fit of the data according to Eq.5 (see text).

Figure 1

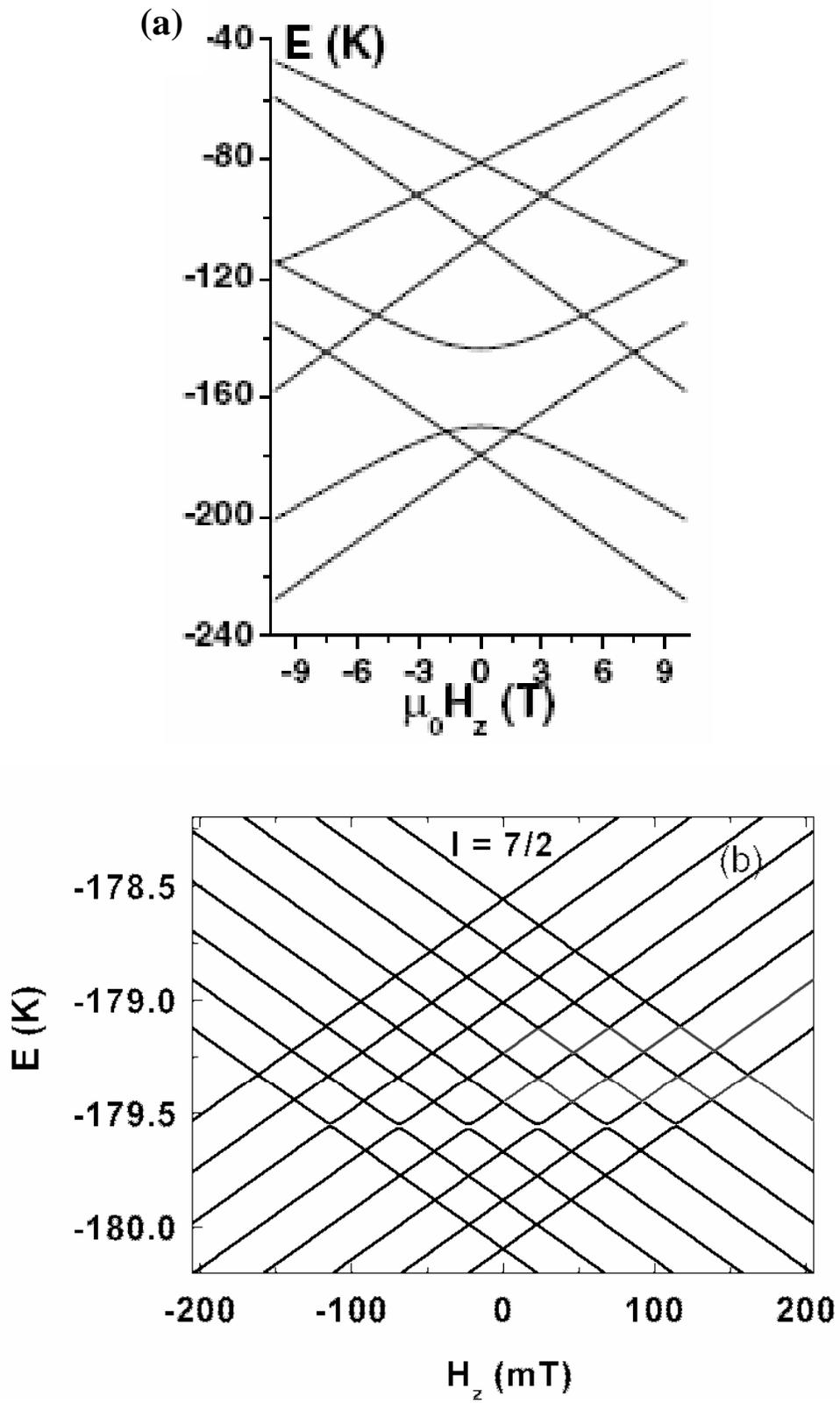

Figure 2

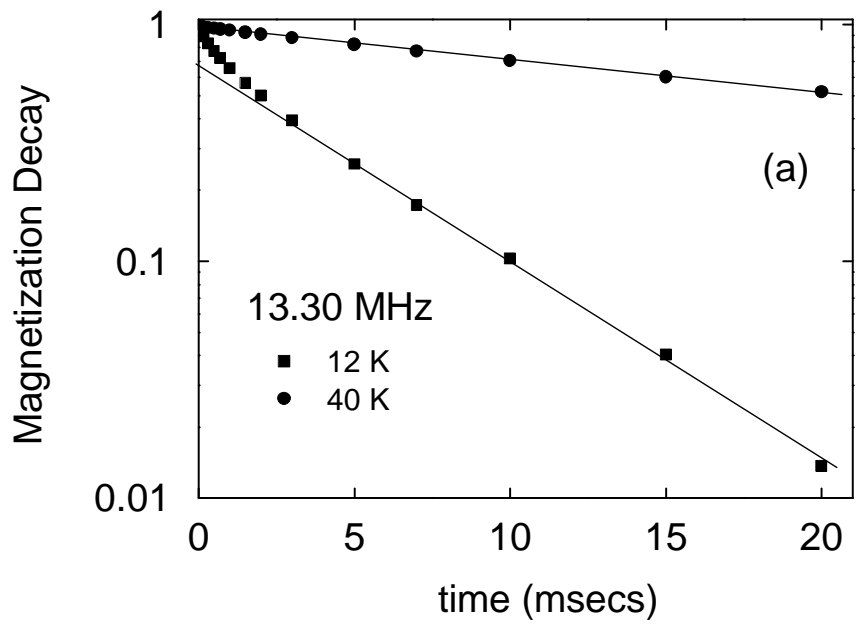

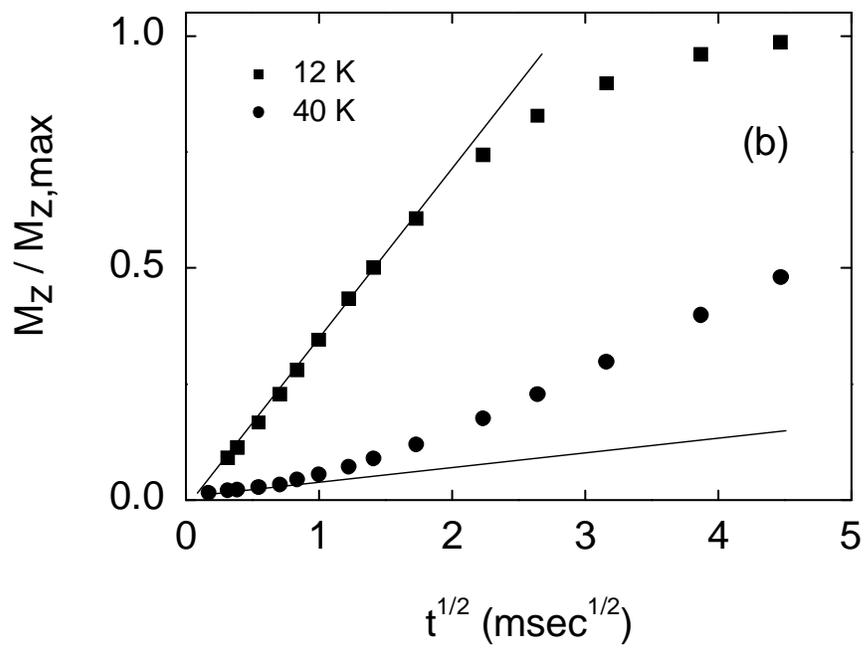

Figure 3

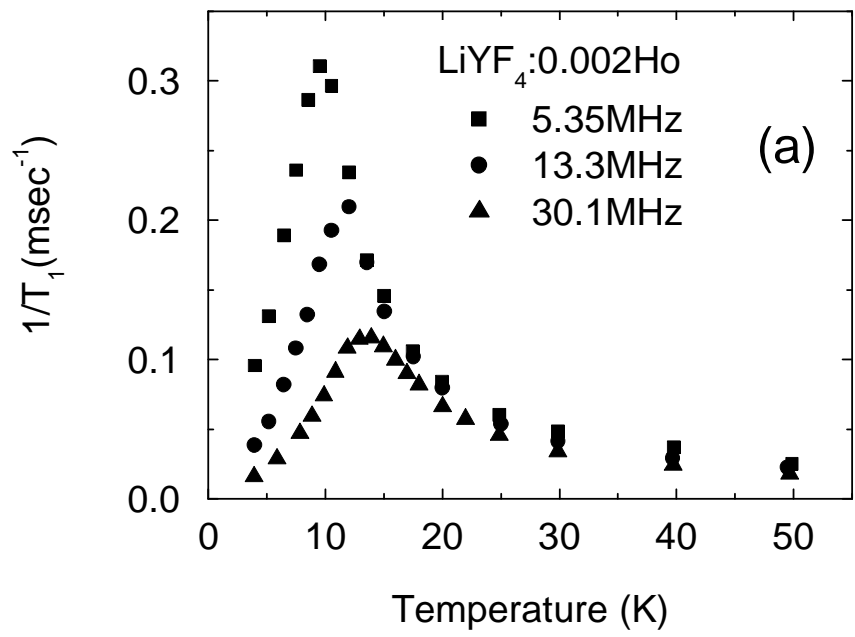

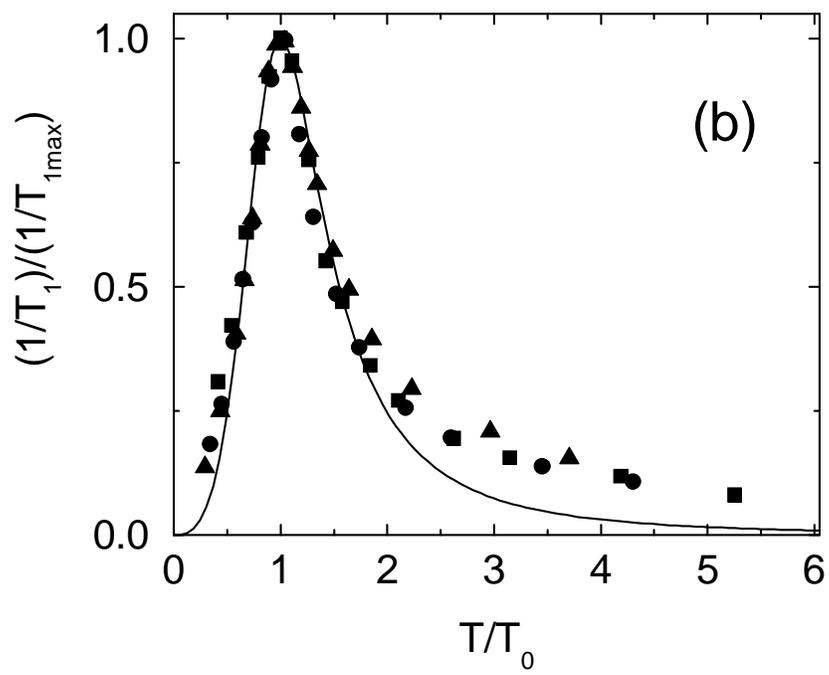

Figure 4

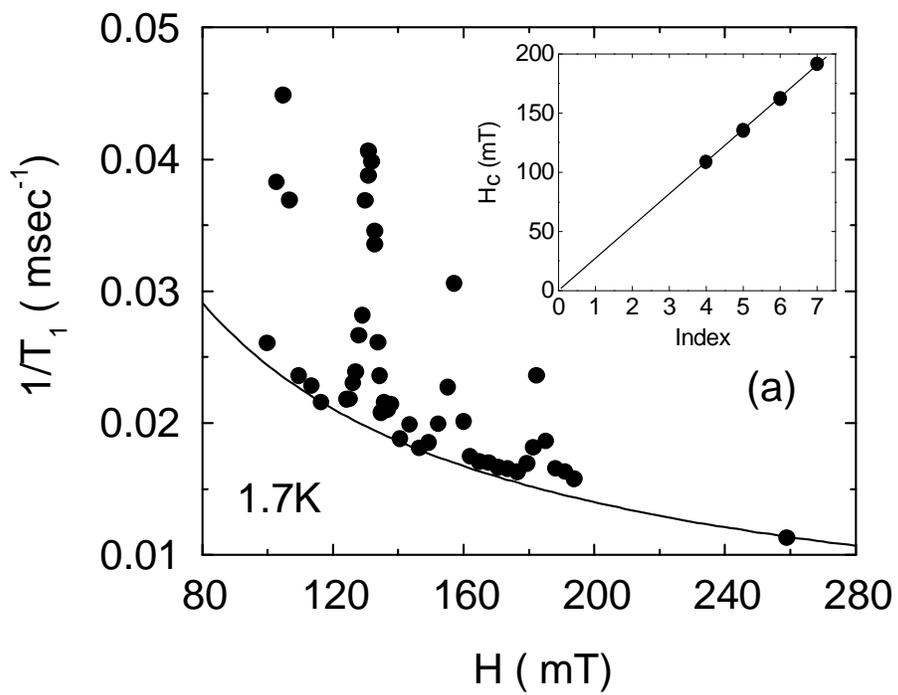

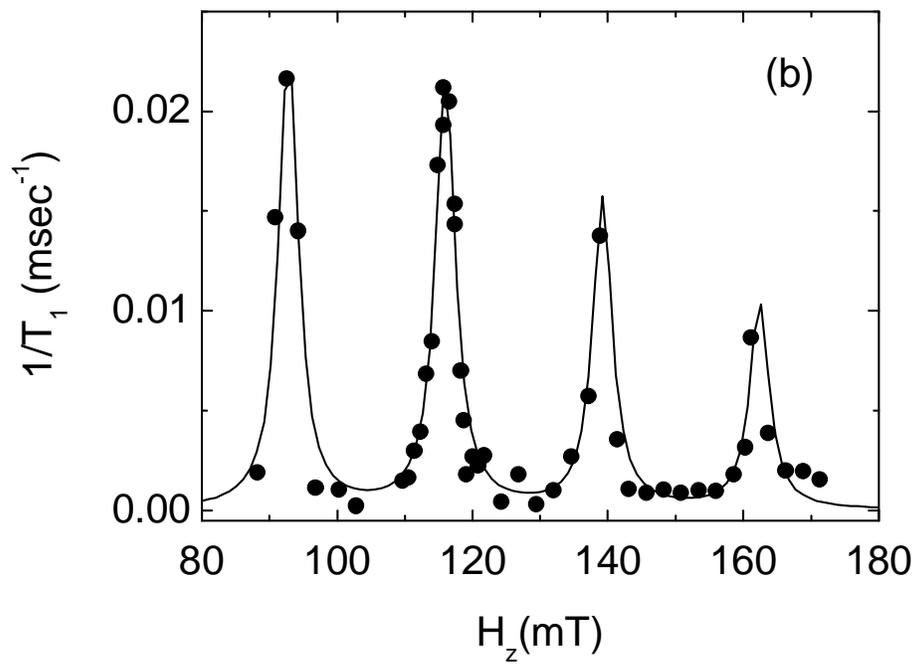

Figure 5

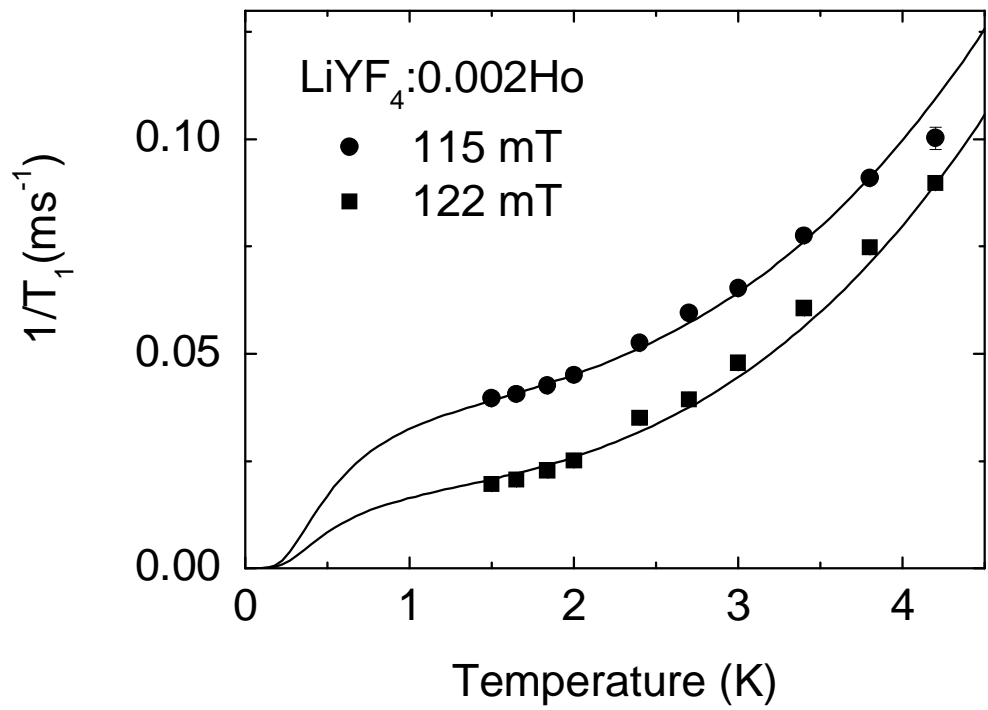

21